\documentstyle[prd,aps]{revtex}

\def\be{\begin{equation}}
\def\ee{\end{equation}}
\def\bea{\begin{eqnarray}}
\def\eea{\end{eqnarray}}
\def\nn{\nonumber}
\def\ep{\epsilon}
\def\c{\cite}
\def\m{\mu}
\def\ga{\gamma}

\def\Ga{\Gamma}

\def\si{\sigma}

\def\pa{\partial}

\def\ov{\over}

\def\rs2r{{r_{s}\over 2r}}
\def\l2r2{{l^{2}\over r^{2}}}

\def\a2{{l^{2}\over a^{2}}}
\def\b2{{l^{2}\over b^{2}}}

\begin{document}

\title{Axial--Vector Torsion and the Teleparallel Kerr Spacetime}

\author{J. G. Pereira, T. Vargas and  C. M. Zhang} 

\address{Instituto de F\'{\i}sica Te\'orica\\ 
Universidade Estadual Paulista \\
Rua Pamplona 145 \\
01405-900\, S\~ao Paulo SP \\ 
Brazil}

\maketitle

\begin{abstract}
In the context of the teleparallel equivalent of general relativity, we obtain
the tetrad and the torsion fields of the stationary axisymmetric Kerr spacetime.
It is shown that, in the slow rotation and weak field approximations, the
axial--vector torsion plays the role of the gravitomagnetic component of the
gravitational field, and is thus the responsible for the Lense--Thirring effect.
\end{abstract}


\section{Introduction}

In a metric--affine theory of gravitation~\c{heh76,heh95}, the
metric and the connection are considered as independent variables, and the underlying
spacetime presents non\-vanishing curvature, torsion and nonmetricity. On the other hand,
in the special case of teleparallel gravity~\cite{hay79,per1,per3,hoi85}, which is
characterized by the vanishing of curvature and nonmetricity, the relevant spacetime is
the Weitzenb\"ock spacetime~\cite{wei}. As is well known, at least in the absence of
spinor fields, the teleparallel gravity is equivalent to general relativity. In order to
better understand this relationship, a study of the teleparallel version of the exact
solutions of general relativity is
indispensable~\c{hm99,yu96,bae88,vla96,ho97,hls,kaw}. For example, in the context of
general relativity, the Kerr solution is an axisymmetric {\it curved} spacetime
produced by a spherically symmetric rotating source. In the context of
teleparallel gravity, this solution might correspond to an axisymmetric {\it
torsionned} spacetime.

We will use the greek alphabet ($\mu$, $\nu$, $\rho$,~$\cdots=1,2,3,4$) to denote
tensor indices, that is, indices related to spacetime. The latin alphabet ($a$,
$b$, $c$,~$\cdots=1,2,3,4$) will be used to denote local Lorentz (or tangent
space) indices, whose associated metric tensor is $\eta_{ab}= \mbox{diag}
(+1, -1, -1, -1)$. Tensor and local Lorentz indices can be changed into each other
with the use of a tetrad field $h^{a}{}_{\mu}$, which satisfies
\be
h^a{}_\mu \, h_a{}^\nu = \delta_\mu{}^\nu ; \quad
h^a{}_\mu \, h_b{}^\mu = \delta^a{}_b .
\label{orto}
\ee
A nontrivial tetrad field can be used to
define the linear Weitzenb\"ock connection 
\be
\Gamma^{\sigma}{}_{\mu \nu} = h_a{}^\sigma \partial_\nu h^a{}_\mu \; ,
\label{car}
\ee 
with respect to which the tetrad is parallel:  
\be {\nabla}_\nu \; h^{a}{}_{\mu}
\equiv \partial_\nu h^{a}{}_{\mu} - \Gamma^{\rho}{}_{\mu \nu}
\, h^{a}{}_{\rho} = 0 \; . 
\label{weitz}
\ee 
The Weitzenb\"ock connection satisfies the relation 
\be
{\Gamma}^{\sigma}{}_{\mu \nu} = {\stackrel{\circ}{\Gamma}}{}^{\sigma}{}_{\mu
\nu} + {K}^{\sigma}{}_{\mu \nu} \; ,
\label{rel} 
\ee
where
\be
{\stackrel{\circ}{\Gamma}}{}^{\sigma}{}_{\mu \nu} = \frac{1}{2}
g^{\sigma \rho} \left[ \partial_{\mu} g_{\rho \nu} + \partial_{\nu}
g_{\rho \mu} - \partial_{\rho} g_{\mu \nu} \right]
\label{lci}
\ee
is the Levi--Civita connection of the metric 
\be
g_{\mu \nu} = \eta_{a b} \; h^a{}_\mu \; h^b{}_\nu \; ,
\label{gmn}
\ee
and 
\be {K}^{\sigma}{}_{\mu \nu} = \frac{1}{2}
\left[ T_{\mu}{}^{\sigma}{}_{\nu} + T_{\nu}{}^{\sigma}{}_{\mu}
- T^{\sigma}{}_{\mu \nu} \right]
\label{conto}
\ee 
is the contorsion tensor, with 
\be
T^\sigma{}_{\mu \nu} =
\Gamma^{\sigma}{}_{\nu \mu} - \Gamma^ {\sigma}{}_{\mu \nu} \;  \label{tor} 
\ee
the torsion of the Weitzenb\"ock connection~\c{ap}.

The torsion tensor can be decomposed into three irreducible parts under the group
of global Lorentz transformations~\c{hay79}. They are the tensor part
\be
t_{\lambda \mu \nu} = \frac{1}{2} \, \left(T_{\lambda \mu \nu} +
T_{\mu\lambda \nu} \right) + \frac{1}{6} \, \left(g_{\nu \lambda} V_\mu +
g_{\nu \mu} V_\lambda \right) - \frac{1}{3} \, g_{\lambda \mu} \, V_\nu \; ,
\label{pt1}
\ee
the vector part,
\be
V_{\mu} =  T^{\nu}{}_{\nu \mu} \; ,
\label{pt2}
\ee
and the axial--vector part
\be
A^{\m} = {1\over 6}\ep^{\m\nu\rho\si}T_{\nu\rho\si} \; .
\label{pt3}
\ee
In terms of these irreducible components, the torsion tensor is written as
\be
T_{\lambda \mu \nu} = \frac{1}{2} \, \left(t_{\lambda \mu \nu} -
t_{\lambda \nu \mu} \right) + \frac{1}{3} \, \left(g_{\lambda \mu} V_\nu -
g_{\lambda \nu} V_\mu \right) + \epsilon_{\lambda \mu \nu \rho} \, A^\rho \; .
\label{deco}
\ee

Relying on the equivalence alluded to above, we are going in this paper to obtain
the teleparallel equivalent of the general relativity Kerr solution. This solution
will be obtained by solving the zero--curvature field equations with torsion. We will
proceed according to the following scheme. In Section~II, we obtain the tetrad
field, the Weitzenb\"ock connection and the irreducible components of the
torsion tensor for the Schwarzschild solution in both isotropic and Schwarzschild
coordinate systems. As expected, due to the spherical symmetry, the axial--vector
torsion vanishes identically. The weak--field limit is then considered, and we show
that the tensor and the vector parts of torsion combine themselves to yield the
Newtonian force. In Section~III, we obtain the tetrad field, the Weitzenb\"ock
connection and the irreducible components of the torsion tensor for the
exact Kerr solution. The slow--rotation and weak--field limits are investigated
in Section~IV, where we show that the axial--vector tensor appears in these limits
as the gravitomagnetic component of the gravitational field. Discussions and
conclusions are presented in Section~V. We use units in which the speed of light
is set equal to unit: $c = 1$.

\section{The Teleparallel Schwarzschild Solution}

In the spherical, static and isotropic coordinate system
$X^{1}= \rho \sin\theta \cos\phi$, $X^{2}= \rho \sin\theta \sin\phi$,
$X^{3}= \rho \cos\theta$, the tetrad components of the Schwarzschild spacetime
can be obtained from the line element
\be
ds^{2}\equiv g_{\mu\nu}dX^{\mu}dX^{\nu}
      = C(\rho)dt^{2}- D(\rho)(d\rho^{2} + \rho^{2}d\Omega^{2}) ,
\label{dsi}
\ee
where
\be
d\Omega^{2}= d\theta^{2} + \sin\theta d\phi^{2} .
\ee
With the subscript $\mu$ denoting the column index, they are given by~\c{hay79} 
\be
h^{a}{}_{\mu} \equiv \pmatrix{
\sqrt{C} & 0 & 0 & 0 \cr
0 &\sqrt{D} &0  &0\cr
0 &0 &\sqrt{D}  &0 \cr
0 &0&0& \sqrt{D}} ,
\ee 
with the inverse
\be
h_{a}{}^{\mu} \equiv \pmatrix{
\sqrt{C^{-1}} & 0 & 0 & 0 \cr
0 &\sqrt{D^{-1}} &0  &0\cr
0 &0 &\sqrt{D^{-1}}  &0 \cr
0 &0&0& \sqrt{D^{-1}}} .
\ee

In a isotropic coordinate system, $C(\rho)$ and $D(\rho)$ are given
respectively by~\cite{mtw}
\be
C(\rho) = \left(1 - {G M \over 2\rho} \right)^{2} \,
\left(1 + {G M \over 2\rho} \right)^{-2}
\ee
and
\be
D(\rho) = \left(1 + {G M \over 2\rho} \right)^{4} \; ,
\ee
with $M$ the gravitational mass of the central source.
The corresponding nonvanishing components of the torsion tensor are
$(i, j, k, \dots = 1, 2, 3)$ 
\be
T^{0}{}_{0i} = H {\pa\rho \over\pa X^{i}} ; \quad
T^{i}{}_{ij} = J {\pa\rho \over\pa X^{j}} ,
\ee
where
\be
H = {1\over 2} \, \frac{d}{d\rho}[\ln C(\rho)] =
{G M \over 2\rho^{2}}(E + F)
\ee
\be
J = {1\over 2} \, \frac{d}{d\rho}[\ln D(\rho)] = - {G M \over \rho^{2}}F \nn \; ,
\ee
with   
\be
E = \left(1 - {G M \over 2\rho} \right)^{-1};\;\;
F = \left(1 + {G M \over 2\rho} \right)^{-1} \; .
\ee
The torsion vector and the axial torsion--vector are, consequently, 
\be
V_{0} = 0;\;\;\; V_{i} = (H + 2J){\pa\rho \over\pa X^{i}}
\ee
and
\be
A^{\mu} = 0 \; .
\ee
   
Now, the Schwarzschild geometry can also be globally represented by the
Schw\-arzs\-child coordinate system $\{x^{\mu}\}= (t, r, \theta, \phi)$, with the
line element in this case given by
\be
ds^{2}= g_{00}dt^{2} + g_{11}dr^{2}- r^{2}d\theta^{2} -
r^{2} \sin^{2}\theta d\phi^{2},
\label{dss}
\ee
where
\be
g_{00} = (-g_{11})^{-1} = 1 - {r_{s}\over r },
\ee
with $r_{s} = 2 G M$ the Schwarzschild radius. Comparing the line elements in the
isotropic and in the Schwarzschild coordinates, given respectively by Eqs.
(\ref{dsi}) and (\ref{dss}), we see that
\be
C(\rho)= g_{00}, \quad \sqrt{D(\rho)} \, \rho = r, \quad
{\partial \rho \over \partial r } = \sqrt{{-g_{11}\over D(\rho)}} .
\ee
Using the general coordinate transformation
\be
h^{a}{}_{\mu} = {\partial X^{\nu'} \over \partial  X^{\mu} }
h^{a}{}_{\nu'} ,
\ee   
where $\{X^{\mu}\}$ and $\{ X^{\nu'}\}$ are respectively the isotropic and
Schwarzschild coordinates, we obtain the tetrad in the Schwarzschild
coordinate system:   
\be
\label{te1}
h^{a}{}_{\mu} \equiv \pmatrix{
\ga_{00} & 0 & 0 & 0 \cr
0 &\ga_{11}{\rm s}\theta {\rm c}\phi &r\;{\rm c}\theta {\rm c}\phi  &
-r\;{\rm s}\theta {\rm s}\phi \cr
0 &\ga_{11}{\rm s}\theta {\rm s}\phi &r\;{\rm c}\theta {\rm s}\phi  &
r\;{\rm s}\theta {\rm c}\phi \cr
0 &\ga_{11} {\rm c}\theta  &- r\;{\rm s}\theta & 0},
\ee
where we have introduced the following notations: $\ga_{00}=\sqrt{g_{00}} $,
$\ga_{ii}=\sqrt{-g_{ii}}$, ${\rm s}{\theta} = \sin \theta$, and
${\rm c}{\theta} = \cos \theta$. Its inverse is
\be
\label{te2}
h_{a}{}^{\mu} \equiv \pmatrix{
\ga_{00}^{-1} & 0 & 0 & 0 \cr
0 &\ga_{11}^{-1} {\rm s}\theta {\rm c}\phi &r^{-1}\;{\rm c}\theta
{\rm c}\phi & -(r\;{\rm s}\theta)^{-1} {\rm s}\phi \cr
0 &\ga_{11}^{-1}{\rm s}\theta {\rm s}\phi &r^{-1}\;{\rm c}\theta
{\rm s}\phi & (r\;{\rm s}\theta)^{-1} {\rm c}\phi \cr
0 &\ga_{11}^{-1}{\rm c}\theta  &- r^{-1}\;{\rm s}\theta & 0}.
\ee
One can easily verify that the relations (\ref{orto}) and (\ref{gmn}) between
$h^{a}{}_{\mu}$ and $h_{a}{}^{\mu}$ are satisfied.

{}From Eqs.(\ref{te1}) and (\ref{te2}), we can now construct the Weitzenb\"ock
connection, whose nonvanishing components are:
\[
\begin{array}{ll}
\Ga^{0}{}_{01} = [\ln \sqrt{g_{00}}]_{,r} & \Ga^{3}{}_{32}=
\Ga^{3}{}_{23} = {\rm cotg} \theta \nn \\
\Ga^{1}{}_{11} = [\ln \sqrt{-g_{11}}]_{,r} &
\Ga^{1}{}_{22} = - {r / \sqrt{-g_{11}}} \nn \\
\Ga^{1}{}_{33} = \Ga^{1}{}_{22} (\sin\theta)^{2} &
\Ga^{2}{}_{33} = - \sin \theta \cos\theta \nn \\
\Ga^{2}{}_{21} = \Ga^{3}{}_{31}={1 / r} &
\Ga^{2}{}_{12} = \Ga^{3}{}_{13}={\sqrt{-g_{11}} / r} \; , \nn 
\end{array}
\]
where a comma followed by a coordinate denotes a derivative in relation to that
coordinate. The corresponding nonvanishing torsion components are:
\bea
T^{0}{}_{01} &=& - [\ln \sqrt{g_{00}}]_{,r} \nn \\
T^{2}{}_{21} &=& - (1 - \sqrt{-g_{11}})/r, \nn \\
T^{3}{}_{31} &=& - (1 - \sqrt{-g_{11}})/r \; . \nn
\eea
Now, as expected, because $A^{\mu}$ represents a deviation from the spherical
symmetry~\c{nit80}, the axial--vector torsion vanishes identically for a
Schwarzschild spacetime:
\[A^{\mu} \equiv 0 \; .
\]
On the other hand, the vector and the tensor parts of torsion are, respectively,
\be
V_{1} = - \left[ (\ln\sqrt{g_{00}})_{,r} +
{2(1 - \sqrt{-g_{11}})\over  r}\right] \; ,
\label{vector}
\ee
and
\bea
t_{001} &=& - \frac{1}{3} (g_{00})_{,r} + \frac{2 g_{00}}{3 r}
\left(1- \sqrt{-g_{11}} \right)
\label{t1} \\
t_{122} &=& - \frac{r}{2} \, \left(1 - \sqrt{-g_{11}} \right) + \frac{r^2}{6}
\left(\frac{(g_{00})_{,r}}{2 g_{00}} + \frac{2 \left(1- \sqrt{-g_{11}}\right)}{r}
\right)
\label{t2} \\
t_{331} &=& - 2 \sin^2 \theta \left[ - \frac{r(1 - \sqrt{-g_{11}})}{2} +
 \frac{r^2}{6} \left(\frac{(g_{00})_{,r}}{2 g_{00}} +
\frac{2 \left(1- \sqrt{-g_{11}}\right)}{r} \right) \right]
\label{t3} \; .
\eea

In the teleparallel description of gravitation, torsion plays the role of the
gravit\-ation\-al force. In fact, a spinless particle submitted to a
gravitational field will obey the force equation~\cite{per1}
\be
\frac{d u_\rho}{d s} - \Gamma_{\mu \rho \nu} \; u^\mu \, u^\nu =
T_{\mu \rho \nu} \; u^\mu \, u^\nu \; .
\label{glofor}
\ee
It is then an easy task to verify that, in the weak--field limit,
the vector and the tensor parts of the Schwarzschild torsion combine themselves to
yield the Newtonian force,
\be
m \; \frac{d \mbox{\boldmath$u$}}{dt} = - \frac{G M m}{r^2} \,
\hat{\mbox{\boldmath$r$}}  \; ,
\label{new}
\ee
where $\mbox{\boldmath$u$} = (u_r, u_\theta, u_\phi)$,
with $\hat{\mbox{\boldmath$r$}}$ the unit vector in the radial direction. 

\section{The Teleparallel Kerr Solution}

The gravitational field of a rotating mass is described by the
axially symmetric stationary Kerr metric~\cite{mtw},
\be
d s^2 = g_{00} dt^2 + g_{11} dr^2 + g_{22} d\theta^2 +
g_{33} d\phi^2 + 2 g_{03} d\phi\; dt , 
\ee
where
\be
g_{00} = 1 - { r_s r \ov \rho^2}; \;\;g_{11} = - { \rho^2 \ov \Delta};\;\; 
g_{22} = - \rho^2
\ee
\be
g_{33} =  - \left( r^2 + a^2 + {r_s r a^2 \ov  \rho^2} \sin^2{\theta} \right)
\sin^2{\theta}
\ee
\be
 g_{03} = g_{30} =  {r_s r a \ov \rho^2 }\sin^2{\theta}  
\ee
with 
\be
\Delta  = r^2 - r_s r + a^2 \;\; {\rm and} \;\;
\rho^2 = r^2 + a^2 \cos^2\theta \; . 
\ee
In these expressions, $a$ is the angular momentum of a gravitational unit
mass source. If $a=0$, the Kerr metric becomes the Schwarzschild metric in
the standard form.

The corresponding Kerr tetrad is
\be
h^{a}{}_{\mu} \equiv \pmatrix{
\ga_{00} & 0 & 0 &\eta \cr 
0 &\ga_{11}{\rm s}\theta {\rm c}\phi &\ga_{22}\;{\rm c}\theta {\rm c}\phi &
-k {\rm s}\phi \cr
0 &\ga_{11}{\rm s}\theta {\rm s}\phi &\ga_{22}\;{\rm c}\theta {\rm s}\phi  &
k \; {\rm c}\phi \cr
0 &\ga_{11}{\rm c}\theta  &- \ga_{22}\;{\rm s}\theta & 0}, 
\label{tek1}
\ee
with its inverse given by
\be
h_{a}{}^{\mu} \equiv \pmatrix{
\ga_{00}^{-1} & 0 & 0 & 0 \cr 
-kg^{03} {\rm s}\phi &\ga_{11}^{-1}{\rm s}\theta {\rm c}\phi 
&\ga_{22}^{-1}\;{\rm c}\theta {\rm c}\phi  &-k^{-1} {\rm s}\phi \cr
kg^{03} \; {\rm c}\phi&\ga_{11}^{-1}{\rm s}\theta {\rm s}\phi 
&\ga_{22}^{-1}\; {\rm c}\theta {\rm s}\phi &k^{-1} \; {\rm c} \phi \cr
0 &\ga_{11}^{-1}{\rm c}\theta &- \ga_{22}^{-1}\;{\rm s}\theta & 0}, 
\label{tek2}
\ee
where
\be
k^2 = \eta^{2} - g_{33}  \;\; {\rm and} \;\;  \eta = { g_{03} \over
\gamma_{00}}.
\ee
One can verify that the relations (\ref{orto}) and (\ref{gmn}) between
$h^{a}{}_{\mu}$ and $h_{a}{}^{\mu}$ are in fact satisfied by the Kerr
tetrad. Moreover, analogously to the Kerr metric, the Kerr
tetrad reduces to the Schwarzschild tetrad for $a=0$. 

The nonvanishing components of the Weitzenb\"ock connection are: 
\[
\begin{array}{ll}
\Ga^{0}{}_{01} = [\ln \sqrt{g_{00}}]_{,r} &
 \Ga^{0}{}_{13}=kg^{03}\ga_{11} \, {\rm s}\theta \nn \\ 
\Ga^{0}{}_{23}=-kg^{03}\ga_{22} \, {\rm c}\theta &
 \Ga^{0}{}_{31}=\eta_{,r} / \ga_{00}-(k^2)_{,r} g^{03}/2 \nn \\
\Ga^{0}{}_{32}=\eta_{,\theta}/\ga_{00}-
   (k^2)_{,\theta} g^{03}/2 & \nn \\
\Ga^{1}{}_{11} = [\ln \sqrt{-g_{11}}]_{,r}  &
 \Ga^{1}{}_{12} = [\ln \sqrt{-g_{11}}]_{,\theta} \nn \\
\Ga^{1}{}_{22} =-\ga_{22}/\ga_{11} &
 \Ga^{1}{}_{33} = -k \, {\rm s} \theta/\ga_{11} \nn \\
\Ga^{2}{}_{12} = \ga_{11}/\ga_{22} &
 \Ga^{2}{}_{21} = [\ln \sqrt{-g_{22}}]_{,r} \nn \\ 
\Ga^{2}{}_{22} = [\ln \sqrt{-g_{22}}]_{,\theta} &
 \Ga^{2}{}_{33} = - k \, {\rm c}\theta/\ga_{22} \nn \\
\Ga^{3}{}_{13} = \ga_{11} \, {\rm s}\theta/k &
 \Ga^{3}{}_{23} = \ga_{22} \, {\rm c}\theta/k \nn \\
\Ga^{3}{}_{31} = [\ln k ]_{,r} &
\Ga^{3}{}_{32} = [\ln k]_{,\theta} \; . \nn
\end{array}
\]
The corresponding non--zero torsion components are:
\bea
T^{0}{}_{01} &=& -[\ln \sqrt{g_{00}}]_{,r} \nn \\
T^{0}{}_{13} &=& \eta_{,r} /\ga_{00} - kg^{03}(k_{,r} -
    \ga_{11} \, {\rm s}\theta) \nn \\
T^{0}{}_{23} &=& \eta_{,\theta}/\ga_{00}-
    kg^{03}(k_{,\theta} - \ga_{22} \, {\rm c}\theta) \nn \\
T^{1}{}_{12} &=& -[\ln \sqrt{-g_{11}}]_{,\theta} \nn \\
T^{2}{}_{12} &=& [\ln \sqrt{-g_{22}}]_{,r} - \ga_{11}/\ga_{22} \nn \\
T^{3}{}_{13} &=& (k_{,r} - \ga_{11} \, {\rm s}\theta)/k \nn \\
T^{3}{}_{23} &=& ( k_{,\theta} - \ga_{22} \, {\rm c}\theta)/k \; . \nn
\eea

The non--zero components of the vector torsion are, consequently,
\be
V_{1} =-[\ln \sqrt{g_{00}}]_{,r} - [\ln \sqrt{-g_{22}}]_{,r}
       + \ga_{11}/\ga_{22} - [\ln k]_{,r} + \ga_{11} {\rm s}\theta/k 
\ee
\be
V_{2} = - [\ln \sqrt{-g_{11}}]_{,\theta} - [\ln k ]_{,\theta} +
       \ga_{22} \, c\theta/k \; ,
\ee
whereas the torsion axial--vector components are
\be
A^{(1)}\times(6h) = -2 (g_{00}T^{0}{}_{23} + g_{03}T^{3}{}_{23}) \nn
\ee
\be
A^{(2)}\times(6h) = 2[g_{00}T^{0}{}_{13} + g_{03}(T^{3}{}_{13} +
T^{0}{}_{01})] \; , \nn
\ee
where we have made the identification
\be
h=\sqrt{-g} \; , \nn
\ee
with $h=\det(h^a{}_\mu)$ and $g=\det (g_{\mu \nu})$.

\section{Slow--Rotation and Weak--Field Approximations}

In the case of slow rotation, we keep the terms up to first order in the
angular momentum $a$. The related quantities are simplified as follows:
\be 
\Delta  = r^2 - r_s r; \quad \rho^2 = r^2 
\label{delta}
\ee
\be
g_{00}=(-g_{11}) ^{-1} = 1 - { r_s  \ov r}; \quad g_{22}= - r^2
\label{g00}
\ee
\be
g_{33}= - r^2 \sin^2{\theta};\quad g_{03} = {r_s a \ov r }\sin^2{\theta} .  
\label{g33}
\ee
In this approximation, both the vector and the tensor parts of torsion reduce to
the values of the Schwarzschild solution. On the other hand, in the weak--field limit,
characterized by keeping terms up to first order in $\alpha$, the nonzero components
of the axial--vector torsion become
\be
A^{(1)}\times(6h) = -2 (\ga_{00} \, \eta)_{,\theta} =
- 2 (g_{03})_{,\theta}
\ee
\be
A^{(2)}\times(6h) = 2 [\ga_{00} \, (\eta)_{,r} - \eta \, (\gamma_{00})_{,r}] ,
\ee
where now
\be
h=r^2 \sin\theta .
\ee
Substituting Eqs.(\ref{g00}) and (\ref{g33}), we obtain 
\be
A^{(1)} = - {2 \ \alpha \ a\ov 3 \ r^2} \, \cos{\theta}
\ee
and
\be
A^{(2)} = - {\alpha a\ov 3 r^{3}} \sin{\theta} ,
\ee
where $\alpha = (r_s/r)$. In spacelike vector form, the axial--vector 
becomes,
\be
\mbox{{\boldmath $A$}} = A^{(1)} \gamma_{11} \; {\bf e}_{r} +
A^{(2)} \gamma_{22} \; {\bf e}_{\theta} ,
\ee
that is,
\be
\mbox{{\boldmath $A$}} = {\alpha \, a\ov 3 r^{2}} [2 \cos{\theta} \; {\bf e}_{r}
+ \sin{\theta} \; {\bf e}_{\theta} ] . 
\label{av1}
\ee

It has been shown by many
authors~\c{hay79,nit80,aud81,yas80,ham95,rum79,heh71,tra72} that the spin
precession of a Dirac particle in torsion gravity is intimately related to
the axial--vector,
\be
\frac{d{\bf s}}{dt} = - {\bf b} \times {\bf s}
\ee
where ${\bf s}$ is the spin vector, and ${\bf b} = 3 \mbox{{\boldmath $A$}}/2$.
Therefore,
\be
{\bf b} = {G J \ov  r^{3}} [2 \cos{\theta} \;
{\bf e}_{r} + \sin{\theta} \; {\bf e}_{\theta} ]
\ee
with $J=Ma$ the angular momentum. Denoting $\mbox{{\boldmath $J$}} = J {\bf e}_{z}$,
this equation can be rewritten in the form
\be
{\bf b} = {G \ov r^{3}} \left[- \mbox{{\boldmath $J$}} +
3 (\mbox{{\boldmath $J$}} \cdot {\bf e}_{r}){\bf e}_{r} \right] \; .
\ee
This means that
\be
{\bf b} = {\bf \Omega}_{LT} \; ,
\label{av2}
\ee
where ${\bf \Omega}_{LT}$ is the Lense--Thirring precession angular velocity,
which in general relativity is produced by the gravitomagnetic component of the
gravitational field~\cite{cw}. We see in this way that the axial--vector torsion
$\mbox{\boldmath$A$}$, in teleparallel gravity, represents the gravitomagnetic
component of the gravitational field. In fact, considering the slow--rotation and
weak--field approximations, the trajectory of a particle, from Eq.(\ref{glofor})
with the Kerr torsion, is found to be
\begin{equation}
m \; \frac{d {\mbox{{\boldmath$u$}}}}{dt} = m \left( - \frac{G M}{r^2} \;
\hat{\mbox{\boldmath$r$}} +
{\mbox{{\boldmath$u$}}} \times {\mbox{{\boldmath$A$}}} \right) \; ,
\label{newlt}
\end{equation}
from where we see clearly that the axial--vector torsion $\mbox{\boldmath$A$}$ is
the gravitomagnetic component of the gravitational field.
  
\section{Final Remarks}

We have obtained in this paper the teleparallel versions of the Schwarzschild
and the stationary axi--symmetric Kerr solutions of general relativity. In the
first case, as expected, due to the spherical symmetry of the Schwarzschild
solution, the axial--vector torsion vanishes identically. We have
then considered the weak--field limit, and we have shown that in this limit the
tensor and the vector parts of torsion combine themselves to yield the Newtonian
force.

In the second case, we have obtained the torsion tensor, as well as the vector and
axial--vector parts of the torsion for the teleparallel Kerr solution. By
considering then the slow--rotation and weak--field approximations, we have shown
that the axial--vector torsion is nothing but the gravitomagnetic
component of the gravitational field, and is therefore the responsible for the
Lense--Thirring effect.

\section*{Acknowledgments}

The authors would like to thank F. W. Hehl for useful comments. They would like
to thank also CNPq--Brazil, CAPES--Brazil and FAPESP--Brazil for
financial support.


\begin{thebibliography}{10}

\bibitem{heh76}
Hehl F W, von der Heyde P, Kerlick G D and Nester J M 1976 Rev. Mod.
Phys. {\bf 48} 393

\bibitem{heh95} 
Hehl F W, McCrea J D, Mielke E W and 
Ne'eman Y 1995 Phys. Rep. {\bf 258} 1 

\bibitem{hay79}
Hayashi K and Shirafuji T 1979 Phys. Rev. D {\bf 19} 3524 

\bibitem{per1}
de Andrade V C and Pereira J G  1998 Phys. Rev. D{\bf 56}, 4689 
de Andrade V C and Pereira J G  1997 Gen. Rel. Grav. {\bf 30} 263 

\bibitem{per3}
de Andrade V C and Pereira J G  1999 Int. J. Mod. Phys. D {\bf 8} 141

\bibitem{hoi85}
M\"uller-Hoissen F and Nitsch J 1985 Gen. Rel. Grav. {\bf 17} 747
 
\bibitem{wei}
Weitzenb\"ock R 1923 {\em Invariantentheorie} 
(Noordhoff, Gronningen)

\bibitem{hm99} 
Hehl F W and  Macias A 1999 Int. J. Mod. Phys. D{\bf 8} 399 

\bibitem{yu96}
Obukhov Yu N, Vlachynsky E J, 
Esser W, Tresguerres R and Hehl F W 1996 Phys. Lett. A {\bf 220} 1

\bibitem{bae88}
Baekler P, G\"urses M, Hehl F W and McCrea J D
1988 Phys. Lett. A {\bf 128} 245 

\bibitem{vla96}
Vlachynsky E J, Tresguerres R, Obukhov Yu N and Hehl F W
1996 Class. Quant. Grav. {\bf 13} 3253 

\bibitem{ho97} 
Ho J K, Chern D C and Nester J M 1997 Chin. J. Phys. {\bf 35} 640

\bibitem{hls}
Hehl F W, Lord E A and Smalley L L 1981
Gen. Rel. Grav. {\bf 13} 1037 

\bibitem{kaw}
Kawai T and Toma N 1992 Prog. Theor. Phys. {\bf 87} 583 

\bibitem{ap}
Aldrovandi R and Pereira J G 1995 {\em An Introduction to Geometrical
Physics} (Singapore: World Scientific)

\bibitem{mtw}
Misner C W, Thorne K S and Wheeler J A 1973
                {\it Gravitation\/}
                (San Francisco: W. H. Freeman and Company);
                Weinberg S W 1972 {\em Gravitation and Cosmology}
                (New York: Wiley);
Ohanian H C and Ruffini R 1994 {\em Gravitation and Spacetime \/}
(New York: Norton \& Company)

\bibitem{nit80}
Nitsch J and Hehl F W  1980 Phys. Lett. B {\bf 90} 98 

\bibitem{aud81}
Audretsch J  1981 Phys. Rev. D {\bf 24} 1470 

\bibitem{yas80}
Yasskin P B  and Stoeger W R 1980 Phys. Rev. D {\bf 21} 2081 

\bibitem{ham95}
Hammond R T 1995 Cont. Phys. {\bf 36} 103  

\bibitem{rum79}
Rumpf H 1980 in {\it Cosmology and Gravitation}, eds. Bergmann P G and 
de Sabbata V (New York: Plenum)

\bibitem{heh71}
Hehl F W 1971  Phys. Lett. A {\bf 36}  225 

\bibitem{tra72}
Trautman A 1972 Bull. Acad. Pol. Sci. Ser. Sci. Math. 
Astron. Phys. {\bf 20}, 895  

\bibitem{cw}
Ciufolini I and Wheeler J A 1995 {\em Gravitation and Inertia}
(Princeton: Princeton University Press)

\end{thebibliography}
\end{document}